\documentstyle[preprint,aps,prd]{revtex}
\draft
\begin{document}
\title{
\begin{flushright}{\normalsize TPI--MINN--95--19/T \\
                               NUC--MINN--95--15/T \\
                               HEP--MINN--95--1344 \\
                               26 July 1995 \\}
\end{flushright}
Magnetic Fields Produced by Phase Transition Bubbles in the
Electroweak Phase Transition}

\author{Gordon Baym}

\address{Department of Physics, University of Illinois at
        Urbana-Champaign, Urbana, IL 61801}
\author{Dietrich B\"odeker and Larry McLerran }

\address{ School of Physics and Astronomy,
	University of Minnesota, Minneapolis, MN 55455}

\date{\today}

\maketitle

\begin{abstract}
    The electroweak phase transition, if proceeding through nucleation
and growth of bubbles, should generate large scale turbulent flow,
which in turn generates magnetic turbulence and hence magnetic fields
on the scale of turbulent flow.  We discuss the seeding of this
turbulent field by the motion of the dipole charge layers in the phase
transition bubble walls, and estimate the strength of the produced
fields.
\end{abstract}

\narrowtext

\section{Introduction}

    The problem of the generation of magnetic fields is an old one in
cosmology \cite{parker,zeldovich}.  The general approach is to
identify mechanisms for the generation of seed fields
\cite{turner}-\cite{olinto}, which can become amplified into fields on
galactic scales.  Considerable attention has been focused on
generation of seed fields during cosmic phase transitions, e.g., Ref.
\cite{hogan}, particularly at the electroweak transition \cite{ew}, or
at the QCD transition through formation of electromagnetic fields in
the collision of shock fronts \cite{qcd} or formation of dipole charge
layers on the surfaces of phase transition bubble walls whose motion
generates an electromagnetic current and hence a magnetic field
\cite{olinto}.  Always present as a source of seed fields are random
magnetic field fluctuations on a scale of order a thermal wavelength.
Once one has identified the source of microscale seed fields, it is
necessary to find how they are transformed into macroscopic scale
magnetic fields, a process requiring production of current loops on
large size scales, and thus the generation of a level of vorticity in
the fluid.

    In this paper we focus on the generation and amplification of
fields at the electroweak phase transition, by means of the fluid, and
hence magnetic, turbulence which should be formed there.  During a
first-order electroweak phase transition bubbles of the new phase form
\cite{linde}-\cite{shaposhnikov} [but see arguments in Ref.
\cite{nigel} why the transition, even if first order transition, need
not necessarily proceed dynamically via domain wall formation].  Then,
as we argue in Sec.\ II, bubble collisions generate a level of
turbulence and hence vorticity in the fluid.  As a bubble propagates
through the electroweak plasma, it generates a precursor shock front
which accelerates the fluid outside the burning front of the bubble.
As the shock fronts from different bubbles collide, they generate
turbulent jets on a scale of order the size of a typical bubble.  The
turbulence in the fluid amplifies whatever seed fields are present to
finite-amplitude large-scale size magnetic fields \cite{landaupk}.
These fields can be considerable, if there is time for equipartition
between the kinetic energy associated with the turbulent flow and the
magnetic field energy.

    A possible mechanism of generation of magnetic seed fields is via
the dipole electromagnetic charge layer on the surfaces of the bubbles
that forms, as discussed in Sec.\ III, as a consequence of baryon
asymmetry and the large mass of the top quark.  The burning surfaces
of bubbles propagating within a region of turbulent flow rotate in the
moving fluid.  The rotation of the dipole charge layer thus sets up a
current in the fluid.  We estimate the typical magnetic field
generated by a single bubble, and then argue that these fields provide
at least one source of seed magnetic fields in the turbulent fluid
which are then amplified into full scale magnetic turbulence.  The
strength of the fields produced this way are considerably smaller in
the absence of magnetic turbulence than the fields that would result
from equipartition in the turbulent flow.

    In order to estimate the generated magnetic field on scale sizes
large compared to the electroweak scale through the cumulative effect
of the many bubbles in the system, we follow the classic analysis of
Hogan \cite{hogan}, and determine, in Sec.\ IV, the magnetic field
correlation function,
\begin{eqnarray}
 C({\bf r}\,) = \left\langle{\bf B}({\bf r}\,){\bf B}(0)\right\rangle.
\end{eqnarray}
If magnetic fields diffuse up to a scale $R$, then the average magnetic
field squared which remains after diffusion will be given in terms of the
correlation function by
\begin{eqnarray}
\left\langle B^2\right\rangle_{R} \sim \left\langle
C({\bf r})\right\rangle_R,
\end{eqnarray}
where $\left\langle \cdots \right\rangle_R$ denotes the spatial
average over a volume of size $\sim R^3$.  From this result we
estimate the typical energy in the magnetic field at any time at
distance scales greater than the magnetic field diffusion length at
that time.  Finally we discuss the possible relevance of this work for
cosmology.  Throughout we use units in which $c=\hbar=1$.

\section{Electroweak phase transition bubbles and turbulence}

    The electroweak phase transition appears, within acceptable
parameters, to be weakly first order \cite{linde}-\cite{shaposhnikov}.
As the universe cools through the electroweak transition temperature,
$T_{\rm c} \sim 100$ GeV, the plasma in the unbroken phase supercools;
eventually, one assumes, small regions of the broken symmetry phase,
with non-zero Higgs expectation value, nucleate, forming bubbles which
expand and fill the system \cite{turok}-\cite{mclerran2}.

    The typical size of a bubble after the phase transition is completed is in
the range
\begin{eqnarray}
R_{\rm bubble} \sim f_{\rm b} H^{-1}_{\rm ew}
\label{bubsize}
\end{eqnarray}
where
\begin{eqnarray}
H^{-1}_{\rm ew} \sim \frac{m_{\rm Pl}}{g_*^{1/2}T_{\rm c}^2}\sim 10{\rm cm}
\end{eqnarray}
is the size of the event horizon at the electroweak scale, $m_{\rm Pl} $
is the Planck mass, $g_*\sim 10^2$ is the number of massless degrees of
freedom in the matter, and the fractional size $f_{\rm b}$ is $\sim
10^{-2}-10^{-3}$ \cite{turok,mclerran1}.  These numbers are typical of
parameters characteristic of the electroweak phase transition.

    The bubble wall surface appears to be stable against small fluctuations
\cite{mclerran2}.  The bubble wall velocity, while poorly known, should be
in the range
\begin{eqnarray}
  v_{\rm wall} \sim 0.05 -0.9.
\end{eqnarray}
When $v_{\rm wall}< 1/\sqrt{3}$, the sound velocity in the symmetric
phase of the electroweak plasma, the burning of the symmetric phase
proceeds by deflagration in which the phase transition burning front
expands out into the symmetric phase.  Ahead of the burning front, a
supersonic shock moves into the symmetric phase, accelerating it
outward, as shown schematically in Fig.\ 1a.  On the other hand, when
the bubble wall velocity is supersonic in the symmetric phase, the
burning discontinuity generates in its wake a similarity rarefaction
wave, with velocity and temperature profiles illustrated in Fig.\ 1b.

    As two shock fronts associated with the nucleation bubbles
collide, they generate turbulence, shown in Fig. 2. At the point where
the shock fronts intersect, the velocity of the fluid from one bubble
is in a different direction than that from the other bubble.  The
geometry is similar to the classic problem of the formation of a
turbulent jet.  The Reynolds number, $Re$, of the flow around two
colliding bubbles is sufficiently large that we expect fully developed
turbulence in a cone associated with the intersection of the two
bubbles, whose opening angle depends on the angle of intersection of
the two fluids \cite{landaufm}.

    The Reynolds number for the collision of two bubbles is
\begin{eqnarray}
   Re \sim \frac{v_{\rm fluid} R_{\rm bubble}}{\lambda},
\end{eqnarray}
where $v_{\rm fluid}$ is the typical fluid velocity, which we take to
be $\sim v_{\rm wall} \sim 10^{-1}$; the typical size of a bubble,
$R_{\rm bubble}$, is given by Eq.\ (\ref{bubsize}).  The typical
scattering length $\lambda$ of excitations in the plasma is of order
\begin{eqnarray}
   \lambda \sim \frac{1}{T g_{\rm ew}\alpha_{\rm w}^2|\ln\alpha_{\rm w}|},
\end{eqnarray}
where $\alpha_{\rm w}$ is the fine structure constant at the
electroweak scale, and $g_{\rm ew}\sim g_*$ is the number of degrees of
freedom that scatter by electroweak processes \cite{baymtrans}.
Therefore
\begin{eqnarray}
   Re \sim 10^{-3} \frac{m_{\rm Pl}}{T_{\rm c}}\alpha_{\rm w}^2
         |\ln\alpha_{\rm w}| \sim 10^{12}
\end{eqnarray}
is sufficiently huge, for any macroscopic size bubble, that the collision
of the shocked matter should generate turbulent flow.

    A turbulent conducting fluid develops magnetic turbulence,
resulting in magnetic fields on all scale sizes.  The relevant time
scale for the amplification of fields on length scale $l$ is of order
$(l/R_{\rm bubble}) t_{\rm trans}$, where $t_{\rm trans}\sim R_{\rm
bubble}/v_{\rm wall}$ is the duration of the phase transition.  If the
field growth is exponential, fields on scales
$l{\mathrel{\raise.3ex\hbox{$<$\kern-.75em\lower1ex\hbox{$\sim$}}}}R_{\rm
bubble}$ can be amplified by many $e$-folds.  When magnetic turbulence
becomes fully developed, the kinetic energy of the turbulent flow is
equipartitioned with that of the magnetic field energy, implying that
the magnetic fields, $B(R_{\rm bubble})$, generated on the scale of
the phase transition bubbles, the largest scale typical of the
turbulent flow, are given by
\begin{eqnarray}
   B^{2}(R_{\rm bubble}) \sim \epsilon(T_{\rm c}) v_{\rm fluid}^2,
\label{equi}
\end{eqnarray}
where $\epsilon(T_{\rm c})\sim g_* T_{\rm c}^4$ is the energy density
of the electroweak plasma.  Since the velocity of the turbulent flow
is $\sim 10^{-1}$, a very large fraction of the energy of the fluid is
in electromagnetic fields.  Turbulent flow is therefore capable of
amplifying any seed field by many orders of magnitude.

    Thermal fluctuations provide a omnipresent source of seed fields
whose size can be estimated by Eqs.  (1) and (2), with the average in
Eq.  (1) the thermal average for a free photon gas.  For scale $R\gg
T_{\rm c}^{-1}$ one has
\begin{eqnarray}
   B^2_{\rm fluct}\sim \frac{T_{\rm c}}{R^3}
   \sim g_*^{3/2} T_{\rm c}^4 f_{\rm b}^{-3}
   \left(\frac{T_{\rm c}}{m_{\rm Pl}}\right)^3
   \left(\frac{R_{\rm bubble}}{R}\right)^3,
\end{eqnarray}
i.e., for $R\sim R_{\rm bubble}$ the value of $B^2_{\rm fluct}$ is
smaller than the equipartition value by a factor of $(g_*^{1/2}/v_{\rm
fluid}^2) f_{\rm b}^{-3} (T_{\rm c}/m_{\rm Pl})^3$.

\section{Magnetic field generated by dipole charge layer in the bubble wall}

    Let us now compare the magnitude of magnetic fields generated by
turbulence with those generated from currents in the bubble walls in
the absence of magnetic turbulence.  These currents arise from the
electric dipole layer that develops in the bubble wall, as a
consequence of the baryon asymmetry of matter undergoing the
transition combined with the fact that the mass of the top quark is
comparable to the phase transition temperature.

    The net baryon number is non-zero in the broken symmetry phase.
If baryogenesis is not associated with the electroweak phase
transition, then the local baryon density in the neighborhood of the
bubble wall is characterized by a baryon number chemical potential
$\mu_b \sim 10^{-9}T_{\rm c}$.  On the other hand, if the baryon
asymmetry is generated at the electroweak phase transition, the
asymmetry is driven by CP violating effects and $\mu_b$ can be much
larger.  We take as an acceptable range $10^{-9} < \mu_b/T_{\rm c}
<10^{-2}$.

    The expectation value of the Higgs field goes from zero to a
finite value as one traverses the bubble wall into the broken symmetry
phase.  Thus the top quark has a non-zero mass inside the bubble, and
zero mass outside.  Top quarks present outside the bubbles face a
potential barrier at the bubble wall, which leads to a Boltzmann
suppression of their number inside the bubble and near the surface.
The slight excess of top over anti-top quarks contributes a net
positive charge outside the bubbles, but the charge excess from the
tops is suppressed inside.  Overall charge neutrality is guaranteed by
the presence of a small electrostatic potential, $A^0$, which is more
negative inside the bubble than outside, attracting light positively
charged particles to the interior.

    The electrostatic potential is determined by
\begin{eqnarray}
       -\nabla^2 A^0 = e \rho_{\rm em} (x),
  \label{poisson}
\end{eqnarray}
where $\rho_{\rm em}$ is the total local electric charge.  Expanding
the distributions of the various charged species to leading order in
$A^0$ and the small non-zero charge, baryon and lepton number chemical
potentials, and then for simplicity expanding in powers of $(m_{\rm
top}/T)^2$, we find
\begin{eqnarray}
   e\rho_{\rm em}+ m_{\rm D}^2 A^0 \sim - e  m_{\rm top}^2 (x)\mu_b,
   \label{debye}
\end{eqnarray}
where $m_{\rm D}$ is the Debye mass for massless leptons and quarks
(including the top).  (We neglect a small correction to the Debye
screening term $\sim m _{\rm top}^2(x)$.)  The Debye screening length
$1/m_{\rm D}$ is of order $1/eT$.  In the expected limit where the
thickness, $L_{\rm wall}$, of the bubble wall is $\gg 1/m_{\rm D}$, we
solve Eqs.\ (\ref{poisson}) and (\ref{debye}) by expanding in powers
of $\nabla^2/m_{\rm D}^2$, and find
\begin{eqnarray}
   e\rho_{\rm em}(x) \sim \frac{e\mu_b}{m^2_{\rm D}}\nabla^2 m_{\rm
top}^2(x)
\end{eqnarray}
plus corrections involving higher order powers of $1/(m_{\rm D}L_{\rm
wall})^2$. The charge density on the surface, $x\approx 0$, is of order
\begin{eqnarray}
   e \rho_{\rm em}(0) \sim\frac1e  \eta_b  T_{\rm c}^3 \varepsilon,
\end{eqnarray}
where the typical baryon asymmetry near the bubble wall is $\eta_b \sim
\mu_b/T_{\rm c}$, $m_D^2\sim e^2 T_{\rm c}^2$, and
\begin{eqnarray}
   \varepsilon \equiv \left( \frac{m_{\rm top}}{T_{\rm c}^2 L_{\rm
   wall}} \right)^2.
\end{eqnarray}
Typically $T_{\rm c}L_{\rm wall} \sim$ 10-100, so that $\varepsilon \sim
10^{-2} - 10^{-4}$.  For $m_{\rm top}(x)$ monotonically decreasing from inside
to out across the bubble surface, we see that the bubble has a dipole charge
layer of order $e\rho_{\rm em}(0)L_{\rm wall}$ per unit area.

    In order for the dipole charge layer on the bubble wall to
generate a magnetic field the bubble must have a net rotation.  If the
bubble is propagating in a turbulent region with vorticity, its
surface acquires a net rotational velocity, $v_{\rm rot}$, of order
the typical turbulent velocity in the fluid, $v_{\rm fluid}\sim
10^{-1}$.  A rotating bubble thus has associated with it a magnetic
moment of order
\begin{eqnarray}
   M_{\rm bubble}&\sim& e\rho_{\rm em} L^2_{\rm wall}
    R_{\rm bubble}^2 v_{\rm rot}\nonumber\\
   &\sim& \eta_b\frac{m_{\rm top}^2}{m_{\rm D}}
   v_{\rm fluid} R_{\rm bubble}^2.
\end{eqnarray}
The magnetic field generated by the bubble on a scale size of order the
bubble radius is therefore
\begin{eqnarray}
   B_{\rm bubble} &\sim& \frac{M_{\rm bubble}}{R_{\rm bubble}^3}
   \sim \eta_b \frac{m_{\rm top}^2 v_{\rm fluid}}
                   {m_{\rm D} R_{\rm bubble}} \nonumber\\
                   &\sim& g_*^{1/2}v_{\rm fluid}
                   T_{\rm c}^2 \frac{\eta_b}{f_{\rm b}}
                   \frac{m_{\rm top}^2}{m_{\rm D} m_{\rm Pl}}.
\label{Bbubb}
\end{eqnarray}
We note that the magnetic field produced by this mechanism is smaller by a
factor $\sim \eta_b T_{\rm c}/m_{\rm Pl }$ compared with the field expected
from equipartition in a magnetically turbulent environment, Eq.\ (\ref{equi}).
At the electroweak scale the equipartition field is $\sim 10^{24}$ gauss, and
the field produced from the dipole layers in the bubble walls, $\sim 10^{-2}$
gauss, is a factor $\sim 10^{-26}$ smaller.

    Cheng and Olinto \cite{olinto}, by contrast, find a field produced by the
dipole layer in the bubbles at the QCD transition, at $T \sim$ 100 MeV, of
order $10^{-10}-10^{-12}$ of the equipartition field, $\sim 10^{18}$ gauss.
The reason for their much larger fraction is that they assume that the width
of the dipole layer is controlled by the width of the baryon diffusion layer,
$\sim 10^{7}$ fm, rather than the microphysics length scale $\sim$ 1 fm at the
QCD scale.  The diffusion layer for the baryon number is so much larger
because the QCD transition occurs sufficiently slowly that the particles have
time to undergo many scatterings during the transition.  In the electroweak
transition on the other hand, the bubble walls move relativistically, and
little diffusion takes place.

\section{Large scale magnetic fields}

    We now ask how magnetic fields on scales larger than the electroweak
bubble size are generated by superposition of the fields associated with the
turbulent flow.  A first estimate can be made by assuming that the magnetic
fields on scales large compared with the size of the turbulent velocity fields
are generated by randomly oriented magnetic dipoles of typical size that of
this velocity field.

    A magnetic dipole $M$ generates a magnetic field far from the dipole of
order
\begin{eqnarray}
  {B}\,(\vec r\,)  \sim e\frac{M}{\mid{\bf r} - {\bf r}_{\rm d} \mid^3 }
   \label{bdip}
\end{eqnarray}
where ${\bf r}_{\rm d}$ is the position of the dipole.  Near the dipole
the field saturates to a more or less constant value.  To estimate the net
magnetic fields produced by randomly oriented dipoles, we make a continuum
approximation for the distribution of dipoles, and assume that the density
$\nu^i(\vec r\,)$ of dipoles pointing in the $i$th direction is Gaussianly
distributed with measure
\begin{eqnarray}
  \int [d\nu^i]\exp\left\{-\frac{1}{2\kappa}\int d^3r\,
  {\vec\nu}^2({\bf r})\right\},
\label{dipdistr}
\end{eqnarray}
where $\kappa$ is a constant.  The correlation function of the density
of dipoles implied by the distribution (\ref{dipdistr}) is
\begin{eqnarray}
    \left\langle\nu^i({\bf r}) \nu^j(0)\right\rangle =
                  \kappa \delta^{ij} \delta^{(3)} ({\bf r}).
  \label{dipcor}
\end{eqnarray}
{}From Eqs.  (\ref{bdip}) and (\ref{dipcor}) we thus find the magnetic
field-magnetic field correlation function
\begin{eqnarray}
   \left\langle {\bf B}({\bf r})\cdot{\bf B}({\bf 0})\right\rangle
   \sim e^2\kappa \int d^3 r_{\rm d}\frac1{\mid{\bf r}-{\bf r}_{\rm
   d}\mid^3} \frac1{\mid{\bf r}_{\rm d}\mid^3}.
\end{eqnarray}
This integral representation is dominated by the regions where either
$\mid{\bf r}_{\rm d}\mid\rightarrow 0$ or $\mid {\bf r}-{\bf r}_{\rm d}\mid
\rightarrow 0$.  The logarithmic divergence of the integral in these regions
is cut off by the size of the typical dipole, $f_{\rm b}H^{-1}_{\rm ew}$, so
that for $r \gg f_{\rm b}H^{-1}_{\rm ew}$,
\begin{eqnarray}
   \left\langle {\bf B}({\bf r})\cdot{\bf B}({\bf 0})\right\rangle
   \sim \frac{e^2\kappa}{r^3}\ln\left(\frac{H_{\rm ew} r}{f_{\rm b}}\right).
\end{eqnarray}

    The average strength of $B^2$ measured by averaging on a size
scale $R$ is thus\footnote{This results differs from Hogan \cite{hogan}, who
on the basis of a random walk of the field lines finds
$\left\langle B^2 \right\rangle_R \sim R^{-3}$.}
\begin{eqnarray}
   \left\langle B^2 \right\rangle_R \sim \frac{e^2 \kappa}{R^3}
   \ln^2\left(\frac{H_{\rm ew}R}{f_{\rm b}}\right).
\end{eqnarray}
If we assume equipartition, and compare with Eq.\ (\ref{equi}), we see
that $\kappa$ is given by
\begin{eqnarray}
   e^2\kappa\sim \left(f_{\rm b} H^{-1}_{\rm ew}\right)^3
   \epsilon v^2_{\rm fluid},
\end{eqnarray}
and therefore
\begin{eqnarray}
   \left\langle B^2 \right\rangle_R\sim v_{\rm fluid}^2 g_* T_{\rm c}^4
    \left( \frac{f_{\rm b}}{H_{\rm ew} R }\right)^3
   \ln^2 \left(\frac{H{\rm ew}R}{f_{\rm b} }\right).
\label{b2}
\end{eqnarray}

    The expansion of the universe causes the magnetic field to decrease as the
square of scale factor of the universe.  Therefore, in the absence of
flux diffusion, the ratio of $\left\langle B^2\right\rangle$ to the energy
$\rho_\gamma$ in photons,
\begin{eqnarray}
         \eta_B \equiv \frac{\left\langle B^2\right\rangle_R} {\rho_\gamma },
\end{eqnarray}
remains independent of time, when measured on a comoving scale size.  From
(\ref{b2}) we have
\begin{eqnarray}
         \eta_B \sim v_{\rm fluid}^2 f_{\rm b}^3
         \left(\lambda_{\rm ew} \over R \right)^3
         \ln^2\left(\frac{R}{f_{\rm b} \lambda_{\rm ew}}\right),
\end{eqnarray}
where $\lambda_{\rm ew}$ is the size of the universe at the electroweak
phase transition times the scale factor, $T_{\rm c}/T_\gamma$, where
$T_\gamma$ is the temperature of the microwave background; at the present
epoch $\lambda_{\rm ew} \sim 10^2$ AU.

    Magnetic fields from the electroweak transition can survive only
on scales on which magnetic diffusion has not had time to wash out the
field correlations.  The flux diffusion equation, which in the local
fluid rest frame can be written as,
\begin{eqnarray}
   \partial_t^2 {\bf B} + k^2 {\bf B} + 4\pi\sigma \partial_t {\bf B}
   = 0,
\end{eqnarray}
where $\sigma$ is the electrical conductivity, implies that a magnetic
field of length scale $1/k$ dies off on a time scale
\begin{eqnarray}
   \tau \sim \sigma /k^2;
\end{eqnarray}
long wavelength magnetic fields set up at the electroweak phase transition
die away very slowly.

    The characteristic diffusion distance at time $\tau$ is therefore
of order
\begin{eqnarray}
   l_{\rm diff} \sim \sqrt{\tau/\sigma};
   \label{ldif}
\end{eqnarray}
at $\tau$ magnetic fields on length scales larger than $l_{\rm diff}$
are not yet significantly affected by diffusion.  An upper limit on
the diffusion length is its present value, about 3 AU \cite{olinto}
which is not much smaller than the size of the horizon at the
electroweak phase transition.

    In the present epoch, $\eta_B\sim (10^{-6} - 10^{-9}) \times
(10^2{\rm AU}/R)^3 \ln^2(R/1{\rm AU})$ for $f_b = 10^{-2} - 10^{-3}$.
Thus the equipartition magnetic field on a scale of order of diffusion
length in the extragalactic medium is now about $B(R\sim10{\rm
AU})\sim 10^{-7} - 10^{-9}$ gauss.  These fields might have acted in
an earlier epoch as seed fields, subsequently ampified by a galactic
dynamo mechanism to produce the galactic magnetic fields.  On a
galactic size scale, at present seven orders of magnitude larger than
the present size of the electroweak horizon, one finds that the
equipartition field is $B(R\sim10^9{\rm AU})\sim 10^{-17} - 10^{-20}$
gauss.

\section*{Acknowledgments}

    We are grateful to Baolian Cheng for stimulating our interest in this
problem, and to Angela Olinto for subsequent discussions.  This research was
supported by the U.S.\ Department of Energy under Grants No.  DOE High Energy
DE-AC02-83ER40105 and No.\ DOE Nuclear DE-FG02-87ER-40328, and NSF Grants PHY
89-21025 and PHY 94-21309.  GB and LM are grateful for the hospitality of the
Los Alamos National Laboratory during the time that this work was initiated,
and of the Aspen Center for Physics where this work was completed.  The work
of DB has been supported by the {\em Deutsche Forschungsgemeinschaft}.

\begin{figure}
\caption{ Bubble structure in (1+1) dimensions for a first-order
electroweak phase transition.  The energy density $\epsilon$ and the
fluid rapidity $\Theta=\ln[(1+v_{\rm fluid})/(1-v_{\rm fluid})]$ are
plotted versus the space-time rapidity $y=\ln[(t+x)/(t-x)]$.  (a)
Deflagration bubble; $\Theta_{\rm def}$ and $\Theta_{\rm sh }$ denote
the rapidities of the deflagration front and the shock front,
respectively.  (b) Detonation bubble; here $\Theta_{\rm s}$ is the
rapidity corresponding to the sound velocity and $\Theta_{\rm det }$
denotes the rapidity of the detonation front.}
\end{figure}

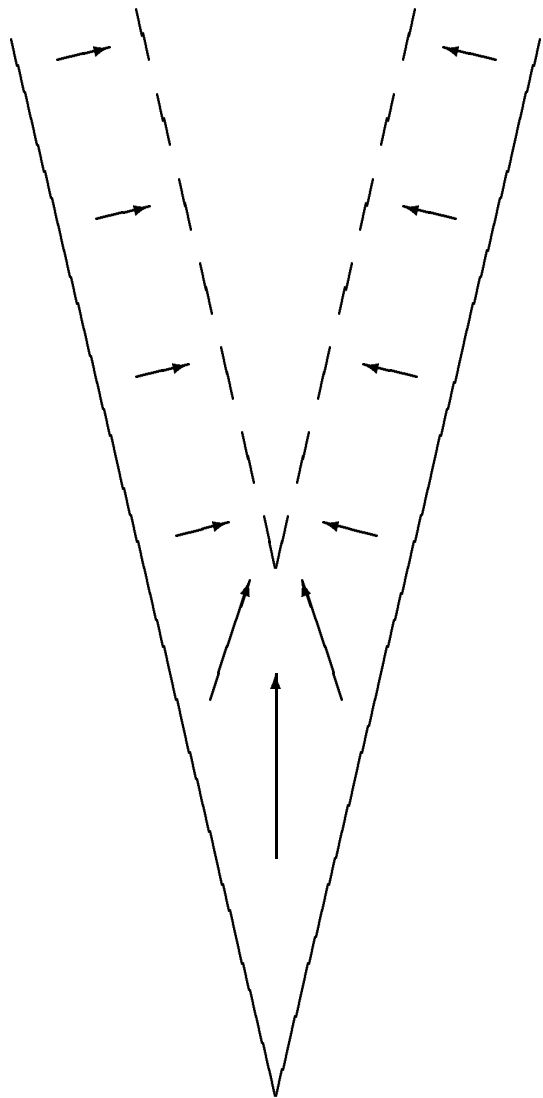
\begin{figure}
\caption{    Fluid velocities during the collision of two bubbles.  The dashed
line represents the shock (detonation) front for a deflagration (detonation)
bubble.}
\end{figure}

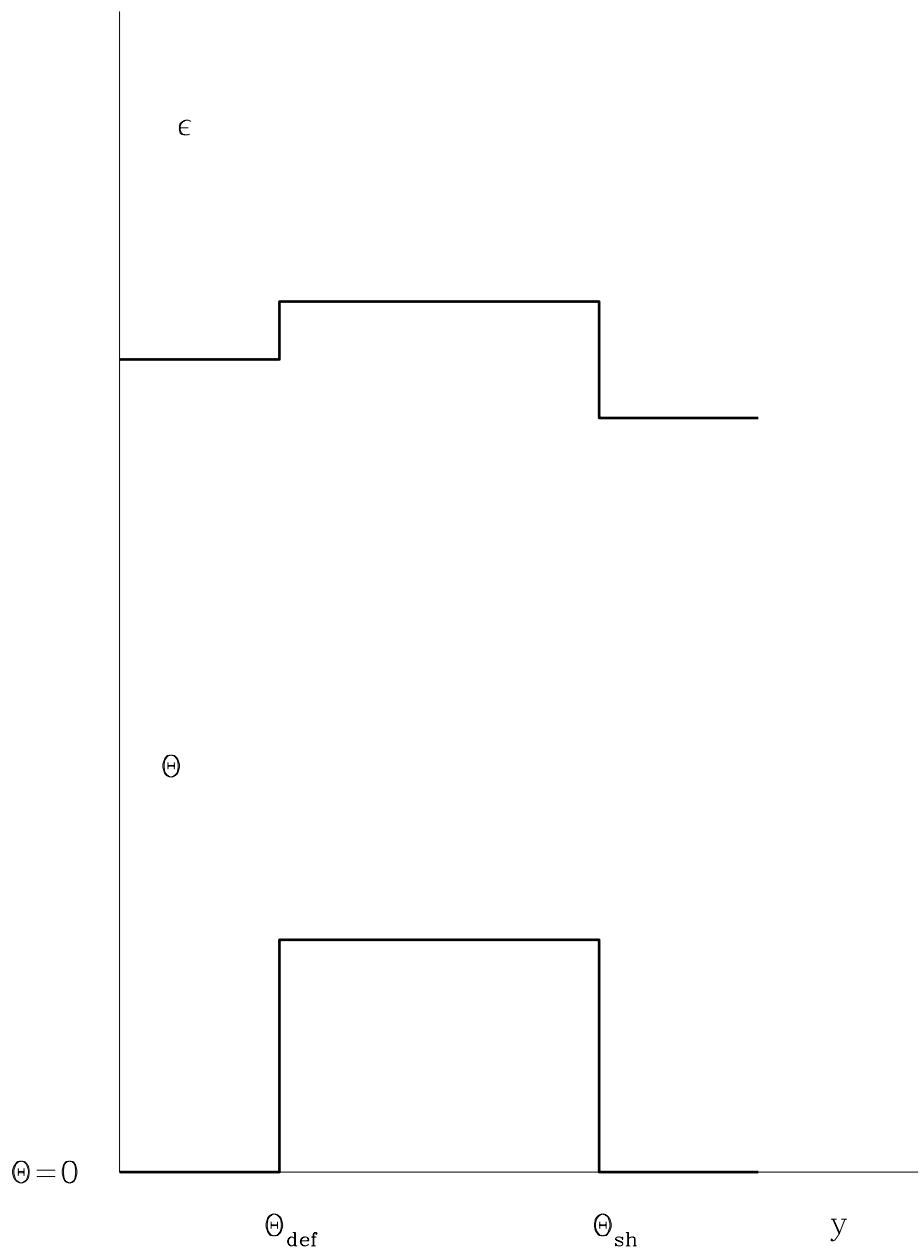
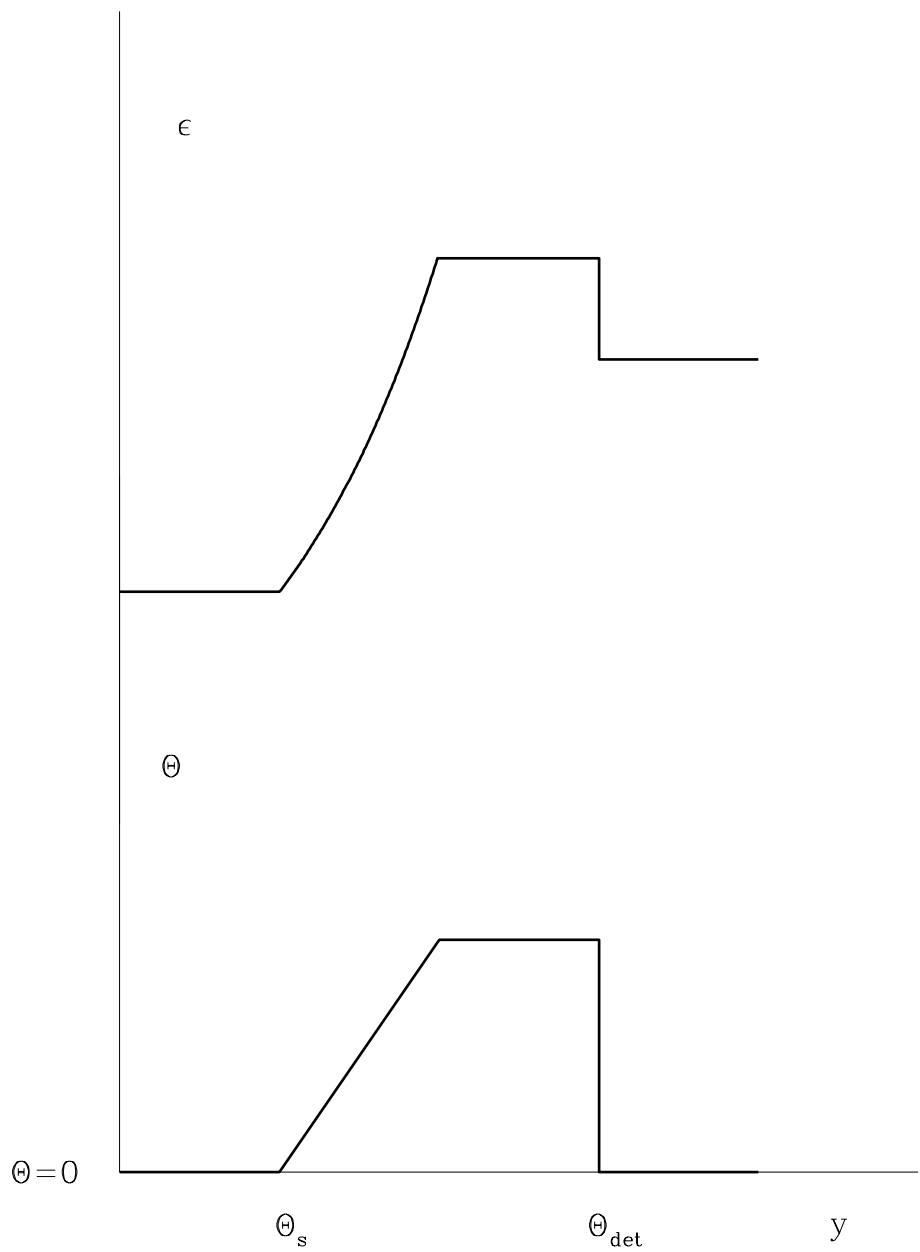
\end{document}